\documentclass[12pt]{article}

\usepackage{amssymb,amsmath,amsfonts,mathrsfs}
\usepackage{hyperref,cite,graphicx}

\allowdisplaybreaks[0]

\makeatletter

\@addtoreset{equation}{section}
\makeatother

\newcommand{\be}{\begin{equation}}
\newcommand{\beq}{\begin{equation}} \newcommand{\eeq}{\end{equation}}
 \newcommand{\ee}{\end{equation}} \newcommand{\bea}{\begin{eqnarray}}
  \newcommand{\eea}{\end{eqnarray}}

\newcommand{\uD}{\mathrm{D}}

\begin{document}

\begin{titlepage}

\setcounter{page}{0}
  
\begin{flushright}
TIFR/TH/10-34
\end{flushright}
  
\vskip 3cm
\begin{center}
  
  {\Large \bf Axions as Quintessence in String Theory }
  
\vskip 2cm
  
{\large  Sudhakar Panda${}^1$, Yoske Sumitomo${}^2$, and Sandip P. Trivedi${}^2$}
 
 \vskip 0.7cm

${}^1$ Harish-Chandra Research Institute,  Allahabad 211019, India\\ 
${}^2$ Tata Institute of Fundamental Research, Mumbai 400005, India
 \vskip 0.4cm

Email: \href{mailto: panda@mri.ernet.in, sumitomo@theory.tifr.res.in, trivedi.sp@gmail.com}{panda@mri.ernet.in, sumitomo@theory.tifr.res.in, trivedi.sp@gmail.com}

\vskip 1.5cm
  
\abstract{\normalsize
 We construct a model of quintessence  in string theory based on the idea of  axion monodromy as
  discussed   by McAllister, Silverstein  and Westphal \cite{McAllister:2008hb}.
  In the model, the quintessence field  is an axion whose shift symmetry is broken
by the presence of 5-branes which are placed in  highly warped throats. This gives rise to a potential for the axion field
which is slowly  varying,  even after incorporating the effects of moduli stabilization and supersymmetry breaking. We find that  the  resulting   time dependence in the   equation of state of
 Dark Energy  is potentially  detectable, depending
on the  initial conditions.  The model has  many very   light extra particles which live 
in the highly warped throats, but  these are hard to detect. A signal in the rotation of the CMB polarization can also possibly arise. 
}
  
\vspace{4cm}
\begin{flushleft}
 November 2010
\end{flushleft}
 
\end{center}
\end{titlepage}

\setcounter{page}{1}
\setcounter{footnote}{0}

\tableofcontents

\parskip=5pt

\newpage

 \section{Introduction}

Observational evidence  shows that the universe is accelerating and  therefore must
be dominated, at cosmological scales, by a mysterious form of energy called Dark Energy. 
Understanding its nature  is a central challenge that faces us today. 
The leading candidate for dark energy is   the cosmological constant. 
It is consistent with observations. Theoretically, we now know that a positive cosmological constant can arise from a 
consistent theory of  quantum gravity like string theory and this gives us greater confidence in the idea being correct. 

Another possibility  for dark energy 
is quintessence, see \cite{Weinberg:2008zzc,Peebles:2002gy,Linder:2007wa,Copeland:2006wr,Tsujikawa:2010sc} for reviews. In this case  the vacuum energy is not a constant but 
instead slowly relaxes due to the evolution of a scalar field. This makes the 
 resulting equation of state for dark energy time dependent. Observations in the coming decade will 
attempt to determine the equation of state  of dark energy and,  hopefully will  allow
 us to decide between these possibilities.

In this paper we will construct a model of quintessence in string theory. 
One can view this as a sort of ``theoretical
test'' of this idea. If the idea fits in with string theory one would have greater confidence in it, just as for 
the cosmological constant. One can also hope that  such  constructions might ultimately  lead to some
 interesting constraints which can then be tested observationally. 

The idea behind quintessence is  quite similar to that in inflation. However, an  important difference, which makes 
the construction of models of quintessence considerably more challenging, is that the energy scale for 
quintessence is of order
\footnote{We will use $\Lambda$ throughout in the following  discussion 
 to   denote this  energy scale. In particular $\Lambda$
 will not denote the cosmological constant in this paper. We hope this does not lead to any confusion.} 
\be
\label{deflambda}
\Lambda \sim 10^{-3} \,{\rm eV}
\ee
 and therefore  is much smaller than the scale of supersymmetry breaking. This leads to two important 
issues  which any model of quintessence must confront. The first is to 
ensure that  the potential for the quintessence field has a scale  of order 
 $\Lambda$  and not of order the SUSY breaking scale; this issue  is related to  the cosmological constant problem. 
The second is to have a  potential for the quintessence field which  meets the slow-roll conditions
despite the relatively high scale of supersymmetry breaking. In particular this requires that the mass of the  
 quintessence field is  of order the Hubble constant today,  $H\sim 10^{-33}\, {\rm eV}$, and therefore is   much smaller than 
the SUSY breaking scale, which must  be at least  
$M_{SB} \sim 1-10 \, {\rm TeV}$ and typically is  much higher \footnote{Here the supersymmetry breaking scale refers to the underlying scale at which the symmetry breaks,  related to the gravitino mass by, $m_{\tilde{g}}\sim M_{SB}^2/M_{Pl}$.}. 
Now, it is a famous fact, responsible for the hierarchy problem in the standard model,  that
 scalar fields tend to have their mass driven up to the SUSY breaking scale. Ensuring that the 
quintessence field has a mass 
$m \sim H$ so that the ratio ${m/ M_{SB}} $ is at least $10^{-45}$ if not  smaller is therefore not easy. 

In this paper we will, to begin,  ignore the  first question, related to the cosmological constant question,
 and focus on the 
 construction  of a model which meets the second challenge of ensuring that the slow-roll conditions are met despite the high scale of SUSY breaking. 
We will return to a discussion of the cosmological constant question in \S 6. Within the context of a model of the 
type we construct, we will argue that anthropic considerations could well result in the energy density in dark energy being 
shared between the cosmological constant and the quintessence field in a reasonably equitable fashion. As a result 
an evolving equation of state might well be more natural from the point of view of anthropic considerations in such a model. 

Before proceeding it is useful to phrase the requirement for a small mass for the quintessence field in terms of restrictions on higher dimensional operators in the low-energy  effective field theory.  
Taking $M_{SB}\sim 1\, {\rm TeV}$, and the mass $m \sim H$, one finds that for the current value of $H \sim 10^{-33} \, {\rm eV},$  
\be
\label{massm}
m^2\sim H^2 \sim {M_{SB}^{8}\over M_{Pl}^{6}}.
\ee
This shows that upto dimension $10$ operators which could contribute to the mass 
 have to be suppressed to construct a satisfactory model.  For a higher scale of SUSY breaking the suppression would have to 
 be even more severe.  The requirements of slow-roll are therefore indeed quite restrictive 
\footnote{The sensitivity
to  UV physics would  be significantly reduced if we were sure that a global symmetry was left intact
by it and only broken at low energies giving rise to the mass. However, in general Planck scale
physics is expected to break global symmetries, so to be sure that the global symmetry is approximately
 unbroken one needs  the full UV theory. We thank L. McAllister for related discussion.}.
 This estimate  
 also illustrates why the construction of  a  quintessence model is sensitive to 
Planck scale physics and  can therefore
 be sensibly carried out only within a UV complete theory of gravity such as string theory. 
In fact in the model we  construct  the canonically normalized quintessence field typically undergoes an excursion 
of order the Planck scale in the course of its evolution and  the potential needs to be flat for this whole range 
 in field space.
Needless to say, ensuring this while working only in an  effective field theory would be extremely difficult. 

Qualitatively, there are two possibilities within string theory for a quintessence field \cite{Witten:2000zk}.
 It could be a modulus like the dilaton
or the overall volume which runs off to infinity in field space. In such a situation the coupling constants of the standard 
model, like
the fine structure constant or the four dim. gravitational constant, would also vary with time and one would also need to 
ensure 
that this variation is small enough. Such models would be observationally very interesting but are quite challenging 
to construct.  Instead, we  will opt here  for a second possibility which seems easier to implement.  
In our  case  the field will be  an  axion, which is a 
sort of Goldstone boson, that runs to a finite 
point in field space where its potential is minimized. We  will take this finite point to be  at the origin.  

It  is well known that   axion fields  often arise in string compactifications. 
 In our construction, which is in IIB string theory, 
 the axion of interest  will arise from the RR sector.
A general argument then ensures that there is a shift symmetry, under which the axion  shifts by a constant, which is unbroken to all
 orders in the quantum loop expansion.  
This feature will be  very helpful in ensuring that the resulting potential is varying slowly enough
and can be isolated from the effects of supersymmetry breaking. 

The shift symmetry will  have  to be broken of course, at some level,  
 to generate a potential for the axion. 
A natural possibility to consider 
 is that the  breaking  arises from non-perturbative effects like instantons.  
 It turns out though 
that  this does not lead to an acceptable  model  for quintessence \footnote{With $O(100)$ axions rolling coherently
 one can get a workable model \cite{Svrcek:2006hf};   at the very least this is an invitation to seek
other possible solution.}. In this case the resulting potential is a periodic function of the axion.
The slow-roll conditions then  require the axion to have a decay constant of order the Planck scale or higher whereas requiring the energy density in the axion to be of the right magnitude gives an axion decay
constant about two orders of magnitude smaller than the Planck scale \cite{Svrcek:2006yi,Svrcek:2006hf}.

One therefore needs  to  explore other ways  to break the shift symmetry.  It turns out that the shift symmetry can 
 also be  broken in the 
presence of branes \footnote{For example, a D$p$ brane in the presence of  the NS two-form $B_2$ acquires an 
induced D$(p-2)$ form charge and 
additional tension. This breaks the $B_2$ shift symmetry although a symmetry still remains under which $B_2$ and the world volume
gauge field $F_2$ both shift keeping $B_2-F_2$ invariant.
   Similarly, the shift symmetry for $C_2$ is broken in the presence of NS5-branes.}. 
 By placing these branes
in highly warped regions of the compactification one can make this breaking small \footnote{At first sight
this way of breaking the shift symmetry might seem rather contrived. But such branes or 
related fluxes can in fact 
arise quite generically in flux compactifications. 
We thank E. Silverstein for a discussion on this point.}. 
This idea is referred to as ``axion monodromy'', and was  developed in 
\cite{Silverstein:2008sg,McAllister:2008hb,Flauger:2009ab} to construct a model of chaotic inflation. 
 The resulting potential one gets in this case
 is no longer periodic, in fact it is approximately linear  in the axion . 
As a result, we will find that the  slow-roll conditions and other requirements for 
 quintessence can be met  even with an axion  whose  decay constant   is 
smaller than the Planck scale.

Following  \cite{McAllister:2008hb} (see also  \cite{Flauger:2009ab}),  we will add  a pair of  NS5-brane and $\overline{\rm NS 5}$-brane
in highly warped regions  of  the 
compactification  to break  the axion shift symmetry.  
 The axion, which is the zero mode of the   RR two-form $C_2$ that arises
due to  a non-trivial two-cycle $\Sigma$, induces a D3-brane charge on the  5-brane which wraps the  two-cycle $\Sigma$. 
It also induces   $\overline{\uD 3}$-brane charge on the anti 5-brane \footnote{The anti 5-brane is   actually
 a 5-brane wrapping cycle $\Sigma$ but with opposite orientation.}. 
This results in additional 3-brane tension being induced on both the 5 and anti 5-branes which depends on the axion. 
The resulting potential turns out  to be  linear in the axion, for large enough values of this field.
 

To examine  the effects of SUSY breaking and moduli stabilization on the axion potential we
 will consider the  model
of KKLT \cite{Kachru:2003aw}.  We will find that an additional important contribution to the axion potential 
does arise from these effects. The axion affects the warp factor of the geometry and this in turn 
changes the volume of the internal space. Once the volume is stabilized this gives rise to an extra potential for the axion. The leading contribution of this type would in fact be unacceptably big, but this 
can cancel exactly between the 5-brane and the anti 5-brane which induce opposite 3-brane charge,
provided the two throats  where they are located are related by a ${\mathbb Z}_2$ symmetry.

The subleading correction to the axion potential which survives cannot be calculated exactly, given 
our present understanding of flux compactifications, but an estimate which is adequate for the purposes
of our discussion can be made following the discussion in  \cite{DeWolfe:2008zy, McGuirk:2009xx, Bena:2009xk}.
This estimate tells us that the  subleading effect is still important enough to be the 
dominant contribution to the axion potential, but it too is suppressed by the warp factor at the
 bottom of the throats where the 5-branes are located \footnote{Although the suppression is by  two powers of the warp 
factor, whereas the 3-brane tension contribution mentioned above is suppressed by  four powers of this factor.}.
By making this warp factor small enough
one then finds that  a satisfactory model of quintessence can indeed be constructed.

The bottom-line then is that  a workable model for quintessence in string theory, based on the idea of axion monodromy,
 can be constructed by placing a 5-brane and anti 5-brane pair in  highly warped throats in the presence of an axion. 
When the canonically normalized axion field has a expectation value which is of order the Planck scale, 
the slow-roll conditions are met, and  the quintessence field evolves with an equation of state which can deviate 
significantly from that for the cosmological constant. 

This paper is structured as follows. Some general considerations for models based on a non-perturbative 
potential  and a   linear potential  are discussed in \S2. 
The quintessence model  we construct   is then discussed in greater detail in \S3.  Additional terms which  can arise once 
the dynamics of moduli stabilization is included are  discussed in \S4.
 \S5 contains a more complete look at the final model including all 
important terms in the potential, and discusses the resulting energy scales in more detail. 
Some general features of the model are discussed in \S6. These include 
 the many very light particles which arise due to the highly warped throats in which the 
5-branes are placed,  the absence of tracker behavior and its relation to the cosmological constant issue,
and a potentially interesting signal due to the rotation of the E mode of the CMB into the B mode.
We end with conclusions in \S7.

Let us end this introduction by commenting on some additional  papers of relevance. 
A model of quintessence based on a linear potential was considered in \cite{Linde:1987},
and was explored further in \cite{Garriga:1999bf,Garriga:2003hj}
\footnote{We thank A. Linde for bringing these references to our attention.}.
The idea that  quintessence might arise from many  axion due to the effects of
flux induced mixing was explored in \cite{Kaloper:2008qs}.  The possibility of many light axions and the constraints that can be placed on them was discussed in \cite{Arvanitaki:2009fg,Arvanitaki:2010sy}. Finally 
\cite{Dong:2010in}, which appeared recently, discusses how additional fields can lead
to flattened potentials.

\section{Axions: General Considerations}
The possibility  of using an axion in string theory as the quintessence field, with the shift symmetry being broken by non-perturbative effects,
  has been considered in the past and has some well known difficulties.  
We begin by reviewing this case first. 

Consider a canonically normalized axion field $\phi$ with action, 
\beq
\label{axact1}
S=\int d^4x \sqrt{-g}\left[-{1\over 2} (\partial \phi)^2 + V\left(\phi\over f_a\right)\right].
\ee
The axion potential in this case is a periodic function with a period of order $f_a$ -  the axion decay constant.
For example, a typical form for the potential is 
\be
\label{perpot}
V=\mu^4 \cos\left({\phi \over f_a}\right),
\ee
where the scale $\mu^4$ goes like, 
\be
\label{scalemuv}
\mu^4\propto e^{-c\, S_{\rm inst}},
\ee
with $S_{\rm inst}$ being   the action of the instanton which gives rise to the potential, and $c$ being a constant of order unity. 

For the axion field to play the role of quintessence its potential must be slowly varying on cosmological scales and therefore
should  meet the well-known slow roll conditions.   
This gives rise   to the requirements that 
\be
\label{slowroll}
{V''\over V}M_{Pl}^2  \lesssim  1, \\ \left({V'\over V}\right)^2 M_{Pl}^2 \lesssim 1.
\ee
For a periodic  potential, e.g., of form eq.(\ref{perpot}), 
 it is easy to see that  $V''/V \sim  ({V'/ V})^2 \sim 1/f_a^2$. So the slow-roll  conditions give
 the constraint 
\be
\label{consta}
f_a \gtrsim M_{Pl}.
\ee

Now  axions which arise in string theory typically  satisfy the condition, 
 \cite{Banks:2003sx,ArkaniHamed:2006dz,Svrcek:2006yi,Svrcek:2006hf},
\be
\label{constb}
f_a\le {M_{Pl}\over S_{\rm inst}},
\ee
where $S_{\rm inst}$ is the action of the  instanton which gives rise to the axion potential.  

For the axion potential $V$ in  eq.(\ref{perpot}), with eq.(\ref{scalemuv}),
 to be of order $\Lambda^4$, eq.(\ref{deflambda}),  requires an action, \cite{Svrcek:2006hf}, 
\be
\label{instact}
S_{\rm inst} \sim  200-300.
\ee

Thus we see that    string constructions lead to an axion decay constant $f_a$, eq.(\ref{constb}),
  which is about two orders of magnitude
smaller than the  value  required by the slow-roll conditions eq.(\ref{consta}).

With $N$ axions evolving together the mismatch does improve. The slow-roll conditions now lead to 
\be
\label{constd}
f_a^2 > {M_{Pl}^2 \over N}.
\ee
However, agreement with eq.(\ref{constb}), eq.(\ref{instact}),  requires, \cite{Svrcek:2006hf},
\be
\label{number}
N \gtrsim 10^4-10^5, 
\ee
which is a large number indeed.  

In the kind of construction we explore, the key new element is that the resulting axion potential is 
not a periodic function. Instead it is   approximately  a {\it linear} function of the axion
\footnote{The more correct form of the potential will be described in later sections. 
The linear approximation is a good one when the axion takes values  in an appropriate range
and otherwise is still a good approximation for making parametric estimates, which is our 
 main purpose  in this section.}
 \cite{McAllister:2008hb}. 
The potential we get has the form 
\be
\label{linpot}
V(\phi)=\mu^4 a = {\mu^4\over f_a} \phi,
\ee
where $\mu$ is a mass scale and we have expressed the
 canonically normalized field $\phi$ in terms of the dimensionless  axion $a$ as
\be
\label{defphi}
\phi=f_a a.
\ee
 
The  resulting action  relevant for the late time evolution of the universe in our model is then
\be
\label{finalaxionact}
S_{axion}=\int d^4x \sqrt{-g}\left[M_{Pl}^2 R - {1\over 2} f_a^2 (\partial a)^2 - \mu^4 a\right].
\ee
 
For  the potential eq.(\ref{linpot}) the slow roll conditions eq.(\ref{slowroll})  give, 
\be
\label{srlinear}
\phi  > M_{Pl}.
\ee
In terms of the dimensionless axion $a$ this condition is, 
\be
\label{srlinear2}
a> {M_{Pl} \over f_a}.
\ee

The second condition we must impose is that the potential energy in  the axion today is 
 of order the total energy 
density in the universe, 
\be
\label{cc}
\mu^4 a \sim \Lambda^4 \sim  10^{-12}\, ({\rm eV})^4
\ee
From eq.(\ref{srlinear2}) this gives rise to a condition on the scale of the potential $\mu$,
\be
\label{condmu}
\mu^4 \lesssim  {f_a\over M_{Pl}} \Lambda^4.
\ee

These are in fact the only two important conditions that need to be imposed. As long as a model can be constructed with a 
linear potential, with a scale $\mu$ which meets the condition eq.(\ref{condmu}), and 
in which the axion around the current epoch of the universe meets the condition in  
eq.(\ref{srlinear2}),  one would have  a workable model of quintessence. 
Note in particular, that the slow-roll conditions, eq.(\ref{srlinear2}),
 can be met in this model even when $f_a\le M_{Pl}$
by choosing a large enough value of the axion. 

Some additional points are also worth making at this stage. 

First, the rolling axion gives rise to a stress energy of the perfect fluid form with an equation of state,
\be
\label{estate}
p=\omega_a \rho
\ee
where 
\be
\label{valomega}
\omega_a=  { {\dot{\phi}^2 / 2} -V \over { \dot{\phi}^2 / 2}+ V} \sim {\epsilon/3-1\over \epsilon/3+1}
\ee
where $\epsilon$ is the slow roll parameter, 
\be
\label{defe}
\epsilon={M_{Pl}^2\over 2} \left({V'\over V}\right)^2 = {M_{Pl}^2 \over 2 \phi^2}.
\ee
When $\phi\sim O(M_{Pl})$ we see that $\omega_a$ can be significantly different from $-1$, 
the value for the equation of state of the  cosmological constant. 

Second, the value obtained for the equation of state parameter 
 (from WMAP+BAO+$H_0$ + $D_{\Delta t} $ +SN) in   \cite{Komatsu:2010fb} is 
\be
\label{boundsexp}
\omega_{\rm DE}(z) = w_0 + w_1 {z \over 1+z}, \quad \omega_0=-0.93\pm 0.13, \quad \omega_1 = -0.41^{+0.72}_{-0.71}.
\ee
Fitting to the central value in our model leads to  the value,  
\be
\label{condaa}
\phi \sim 2.14  M_{Pl}.
\ee
The time-dependent constraint $\omega_1$ is also satisfied.

Third,  the axion satisfies the equation
\be
\label{eqaxion}
\ddot{\phi}+3 H \dot{\phi}+V'=0.
\ee
When the slow roll conditions are satisfied, 
\be
\label{approx}
3 H \dot{\phi}+V'=0,
\ee
the total change in  the axion field  during the evolution of the universe, upto the current epoch, can  be estimated as
\be
\label{changea}
\delta \phi\sim \dot{\phi} H^{-1} \sim {V'\over V}M_{Pl}^2 \sim {M_{Pl}^2 \over \phi}.
\ee
For  $\phi$ of order the Planck scale
 we see that the total change in the axion  during the evolution of the universe 
is also of order the Planck scale.

For a workable  model, the  potential must be slowly varying for  this entire range of variation
of the axion field. In particular, in our case the linear approximation with a slow enough rate of variation,  
 should  be valid for the entire range of evolution of the axion. 
This is a significant constraint  to meet. 
As was discussed in the introduction it would be difficult to ensure this while 
working within a low-energy effective field theory alone. 
 For 
once the axion shift 
symmetry is broken  one would expect  quadratic and higher terms  to typically appear
\be
\label{higherterms}
V={\mu^4 \over f_a} \phi + m^2 \phi^2 + \cdots {\phi^{n+4}\over M_{Pl}^n}
\ee
thereby ruining the flatness conditions required for slow-roll behavior. 
By embedding this construction in string theory 
we can go beyond general field theory considerations though  and as we discuss below, 
 we will 
find that in some  models these corrections do not arise  and  the potential remains of 
linear form, with small corrections, for the entire range of  $O(M_{Pl})$ variation in the canonically normalized axion.

\section{ More Details on the Model}

We now turn to a more detailed description  of the model. 

We will work  with 
flux compactifications  in IIB string theory. This is a reasonably well studied corner of string theory  by now, 
e.g, see \cite{Giddings:2001yu,Douglas:2006es}. 
One starts with a Calabi-Yau manifold and  carries out a  suitable orientifolding to allow for the presence of flux.
Then turning on flux and/or adding branes gives rise to a warped Calabi-Yau internal space.  

The axion field we will consider  arises in this setup  from the zero-mode of a two-form field. 
There are two-possibilities, the NS two-form, $B_2$, or the RR two-form  $C_2$. 
In the   $C_2$ case  we start with  the ten-dim. action, 
\be
\label{tendimact}
S= {1\over (2\pi)^7 \alpha'^4} \int d^{10}x \sqrt{-g} \left[ {1\over g_s^2} R - {1\over 12}   \partial_\mu C_{ab} \partial^\mu 
C_{a'b'} g^{aa'}g^{bb'} + \cdots \right]
\ee
where $\mu = 0, 1 \cdots 3$ labels non-compact directions and $a,b,a'b'$ label the six compact directions. We then 
 reduce to get the four-dim. action. 
 
This gives,  
\be
\label{fourdimact}
S= \int d^4x\sqrt{-g_4} \left[{M_{Pl}^2 \over 2}  R  - f_a^2 {1\over 2} (\partial a)^2\right]
\ee
where  the four-dim. Planck scale is, 
\be
\label{mpl}
M_{Pl}^2={2 L^6\over (2\pi)^7 g_s^2 \alpha'}
\ee
and  the volume of the internal space is 
\be
\label{intvol}
V=L^6 (\alpha')^3,
\ee
with $L$ being the dimensionless modulus. 
Keeping only the dependence on the overall volume  modulus and the dilaton in the axion kinetic energy terms we get, 
\be
\label{deffa}
f_a^2 \sim {g_s^2 M_{Pl}^2  \over  L^4}. 
\ee
As a simple model we can take the internal space to be a six-torus with equal size circles. The axion comes from the zero mode
of say the $T^2$ spanned by the first two internal directions.  This gives, 
\be
\label{axionc2}
a = C_{12}
\ee
with,
\be
\label{defaa}
f_a^2 = {g_s^2 M_{Pl}^2  \over 6 L^4}.
\ee

In the more general situation of a Calabi-Yau orientifold with several moduli,  other moduli will also enter in determining 
$f_a$ but the dependence 
 on the overall volume  and $g_s$ should still be parametrically of the form eq.(\ref{deffa}). 
Since the volume modulus in particular gives
some of the more stringent constraints we will neglect these additional   moduli in our discussion below. 
  A more complete analysis will
 have to include    them of course.  The case  where  all the complex structure moduli 
are much heavier and decouple due to a tree-level superpotential, and the  Calabi-Yau has   only one K\"ahler moduli, will  
essentially map to the case above \footnote{One situation which can  be qualitatively different is if the axion arises from a two-cycle which is localized in a highly warped region.
In this case $f_a$ is suppressed and goes like,  $f_a\sim e^{A_{top}}$, which is roughly the
 maximum value of the warp factor along the localized 2-cycle. As a result meeting the slow-roll conditions requires large values of $a$.  We do not  pursue this possibility  any further below.}.

In the discussion above  we have  considered the case of a RR axion. For  an NS  axion  
the only difference is that the   factor of  $g_s^2$ on the RHS will be missing in  
eq.(\ref{deffa}). As we will see below, it will be advantageous for our purposes to take the axion to arise from the RR field. 
For this reason we  mostly present the formulae for the RR case in our discussion.   

We see  from eq.(\ref{deffa}) that for $g_s<1,\, L>1$,\,  $f_a<M_{Pl}$, in keeping with the general expectations in string theory discussed above \footnote{More precisely, an instanton contribution
 which  breaks the shift symmetry for $a$ would arise from a  ED1 brane wrapping the $T^2$ spanning the first two 
directions. This would have action $S_{\rm inst}\sim L^2/g_s$, so that $f_a\sim M_{Pl}/S_{\rm inst}$ which agrees with 
eq.(\ref{constb}).}. To meet the slow roll conditions for a linear potential we saw above that 
eq.(\ref{srlinear2}) needs to be met. From eq.(\ref{deffa}) we see that  the axion has to satisfy the 
requirement,  
\be
\label{rangeaxion}
a>{L^2 \over g_s}.
\ee

It is well known that the NS and RR  axions have an approximate
 shift symmetry in string theory. For example the NS two-form has a coupling on the world-sheet,
\be
\label{wscoupling}
S_{w.s.}= \int d^2 \sigma B_{AB} {dX^A \over d\sigma^i} \wedge {dX^B \over d \sigma^j} \epsilon^{ij},
\ee
where $A=0, 1, \dots 9$ denote all space-time directions. 
Now for a two-form, $B_{AB}$,  which has   no field strength,  $H=dB=0$,  
 it is easy to see that the right hand side in eq.(\ref{wscoupling}) is  a total derivative and 
must vanish in the absence of any boundary for the world sheet. 
In this way we see that there cannot be a potential which arises for an axion coming from the zero-mode of $B_2$. 
Similarly, for the RR fields it is well known that the vertex operator corresponds to the field strength
 rather than the gauge potential, thereby again leading to no potential. 

The argument in the previous paragraph is   in fact true to 
 all orders in the $\alpha'$ and the $g_s$ expansion. For the NS case they fail once non-perturbative corrections in 
$\alpha'$ are introduced and  world-sheet instantons can give rise to corrections which generate a potential for the
 axion. For the RR case space-time instantons carrying the charges of the relevant Euclidean D1 brane are needed. 
The argument can also fail in the presence of branes. In the presence of a D$p$-brane for example, it is well known that  
a constant $B_2$ field \footnote{More correctly one means $B-F$ where $F$ is the world volume gauge field.}
leads to additional  charges for the brane  (corresponding to  D$p$-branes with smaller $p$) and 
correspondingly additional tension. This happens in particular for a D5 brane, which in the presence of non-zero $B_2$ can 
acquire D3-brane charge. By S-duality, and this will be particularly relevant for our model, it follows then that 
this can also happen for an NS 5-brane in the presence of a $C_2$ field. 

 Our basic strategy will be to break the  approximate shift symmetry, which prevents a potential for the axion, 
 by including branes in a 
highly warped region of the compactification. This will lead to a potential, but one which is suppressed by the warping.

\subsection{The Basic Setup}

Before discussing the breaking of the shift symmetry in more detail let us give some more details about the basic setup
of the model. 
This consists of  a Calabi-Yau manifold in Type IIB string theory with three-form and five-form
fluxes turned on \cite{Giddings:2001yu}. An orientifold projection is needed to be able to turn on the fluxes, more generally one works 
with an F-theory compactification. The orientifold
projection retains states invariant under $(-1)^{F_L}\Omega \sigma$, where $\sigma$ is a ${\mathbb Z}_2$ symmetry of 
the Calabi-Yau manifold. The K\"ahler moduli arise from non-trivial two-cycles.
If there are 2 non-trivial two-cycles $\Sigma_1, \Sigma_2$ then there are correspondingly two K\"ahler moduli in the 
parent Calabi-Yau. Now if these two-cycles are  exchanged by $\sigma$, then after orientifolding only one K\"ahler modulus,
corresponding to the even combination, survives, see \cite{Grimm:2004uq} for more details. 
 Similarly zero modes, related to the 2 non-trivial two-cycles,
 arise for the $B_2,C_2$ two-forms and the $C_4$ four- form.
For $C_4$, after orientifolding the even combination survives, while for $B_2, C_2$ the odd combination survives. 
The orientifold symmetry breaks the supersymmetry down to ${\cal N}=1$. The  zero mode for $C_4$ and  the K\"ahler modulus
which both arise from the even combination of the 2 two-cycles give rise to one chiral superfield which we denote by $T$. 
The zero modes from $B_2,C_2$, denoted by $b,a,$ respectively,  which arise from the odd combination give rise to another chiral superfield  denoted by 
$G=b- \tau a$, where $\tau=C_0+i/g_s$ is the dilaton axion field.   

In general there will be several $T$ and $G$ moduli.
In the discussion below for clarity, we will  simplify and only consider the case where there is one pair of two-cycles
and therefore  one resulting  $T$ and $G$ moduli each.
In this case the real part of $T$ is related to the overall volume modulus eq.(\ref{intvol})  by
\be
\label{repT}
{\rm Re} (T)\sim L^4.
\ee

Besides the various moduli mentioned above additional ones arise from complex structure deformations as well. 
Once the fluxes are turned on these complex structure  moduli along with the dilaton-axion will acquire a mass at tree-level \cite{Giddings:2001yu, Gukov:1999ya}.    
In the KKLT model  after integrating out these moduli there is a resulting superpotential of the form, 
\be
\label{reskklt}
W=W_0+Ae^{-\alpha T}
\ee
which will stabilize the overall volume and the $C_4$ axion. 

SUSY breaking in the KKLT scenario occurs by adding $\overline{\uD 3}$-branes.
Let us denote by $U_{\rm mod}$ the scalar potential which is generated
for moduli stabilization. Then the SUSY breaking scale $M_{SB}$ in the KKLT model is  of order,
\be
\label{umod}
 M_{SB}^4 \sim U_{\rm mod}.
\ee

\subsection{Breaking The Shift Symmetry}
We now turn to breaking the axion shift symmetry.
In our model the  parent 
Calabi Yau manifold has two  ``throats'', or highly warped regions. Later on we will see that it is safest
to have a ${\mathbb Z}_2$ discrete symmetry ${\cal R}$,  present under which the manifold is invariant and  which exchanges the two throats.
This ${\mathbb Z}_2$ symmetry is in addition to the  symmetry $\sigma$ involved in the 
orientifolding, and acts non-trivially on the Calabi-Yau orientifold.

The  two-cycles $\Sigma_1, \Sigma_2$,  descends into each of the two throat. 
A 5-brane  is located at the bottom of the  first throat, where the warp factor acquires its minimum value, 
 and  wraps a combination of the  $2$  two-cycles  that is  invariant under the orientifold symmetry.
In the presence of a non-zero value for  the axion the 5-brane acquires charge,
and additional tension,  
corresponding to  D3-branes which are filling all the non-compact directions and are point-like in the internal 
space 
In addition,  an anti 5-brane is located at the bottom of the second throat and again wraps  an orientifolding-invariant
 combination of the two -cycles. The anti 5-brane can be thought of the 5-brane wrapping the two-cycles with opposite
 orientation. As a result $\overline{\uD 3}$-brane charge is induced on the anti 5-brane in the presence  of the axion
and correspondingly it acquires additional 3-brane  tension. The net configuration is not supersymmetric. 
For additional details see also \cite{McAllister:2008hb}.

Note that the presence of the anti 5-brane  is important for charge cancellation. 
The total D3-brane charge must cancel by Gauss's law. Without
the anti brane the additional 3-brane charge induced  on the 5-brane  would either not have been allowed at all
 or at least  would not be allowed to relax with time. 
 
For a D5-brane in the presence of $B_2$ the D3 charge arises due to the WZ term on its world volume, 
\be
\label{wzd5}
S_{WZD5}=\int B_2 \wedge C_4
\ee
and the additional   tension comes   from the DBI term,
\be
\label{dbid5}
S_{DBID5}= -{1\over (2\pi)^5 g_s\alpha'^3}\int d^6x\sqrt{-\det(g+ B_2)}
\ee
For an NS 5-brane, it then follows by S-duality that there is a WZ coupling  and DBI term involving $C_2$
\be
\label{wzns}
S_{WZNS5}=\int C_2\wedge C_4
\ee
\be
\label{dbins}
S_{NS5}= -{1\over (2\pi)^5 g^2_s\alpha'^3} \int d^6x \sqrt{-\det (g + g_s C_2)}.
\ee

In the latter case this gives rise to a  potential for the axion, 
\be
\label{tension}
V=  2 \epsilon{1\over (2\pi)^5 g_s^2\alpha'^2} \sqrt{ L^4 + g_s^2 a^2}. 
\ee
Here,
$a\alpha' = \int_\Sigma C_2$.  The factor of $2$ is because of including both the 5-brane and anti brane. And as in our discussion of the kinetic energy for the axion above, we 
have only shown the parametric dependence on the overall volume modulus and suppressed the dependence on other moduli.

The parameter $\epsilon$ in eq.(\ref{tension}) arises from the warped geometry. 
Consider   a warped compactification with metric,
\be
\label{metricw}
ds^2=e^{2A(y)} dx_\mu dx^\mu+e^{-2A(y)} g_{ab}dy^ady^b
\ee
where the 
warp factor  at the  location of the 5-brane is $e^{A_0}$. Then $\epsilon$ is given by  
\be
\label{defeps}
\epsilon= e^{4A_0}. 
\ee

It is useful to consider  the limit where the axion has a large  value,  
\be
\label{lva}
a\gg {L^2 \over g_s}.
\ee
In this case the potential becomes linear \footnote{We will take the axion $a>0$, for simplicity.} 
\be
\label{v0}
V=2 \epsilon {1\over (2\pi)^5 g_s\alpha'^2} a. 
\ee
Comparing with eq.(\ref{linpot}) we see that 
\be
\label{valmu}
\mu^4=\epsilon {2\over (2\pi)^5 g_s\alpha'^2}
\ee
and eq.(\ref{defeps}) implies that 
\be
\label{depmu}
\mu^4 \propto e^{4A_0}.
\ee
When eq.(\ref{lva}) is met  the 5-brane tension itself is insignificant compared to the contribution due to the 
3-brane charge. And one can think of the setup as essentially consisting of 
  a  stack of 3-branes in one  throat with 
another    stack of anti 3-branes in the ${\mathbb Z}_2$ image throat.  
 As the axion evolves the induces 3-brane charge decreases and along with it the induced tension of the 3-branes also decreases. 

The linear potential eq.(\ref{v0}) is a good approximation when $a\gg L^2/g_s$, but as noted above the slow-roll conditions
only require that $a \gtrsim  L^2/g_s$. When this later 
more general condition is met, we must use the full form of the potential
in eq.(\ref{tension}). Our analysis will be mostly parametric in nature and to simplify the algebra we will mostly use the linear form below. 

Two conditions need to be met for a workable model, these are summarized in eq.(\ref{srlinear2}) and eq.(\ref{condmu}).  
In the limit, eq.(\ref{lva}) we see from  eq.(\ref{deffa}) that the slow-roll condition eq.(\ref{srlinear2}) is met.  
In addition if the warp factor is small enough 
the axion energy density
\be
\label{ccb}
\rho \sim  \epsilon{2\over (2\pi)^5 g_s\alpha'^2}  a \sim \Lambda^4
\ee
can also be of the required small value.

To investigate this last condition,  let us use eq.(\ref{mpl}) and  express 
the string scale $\alpha'^2$ in terms of $M_{Pl}$ 
and the moduli $L,g_s$. This gives
\be
\label{cepsb}
\epsilon \sim {2 L^{12} \over (2\pi)^9  a g_s^3} \left({\Lambda \over M_{Pl}}\right)^4
\ee
Taking  $L=10, g_s=1$ and $a \sim L^2$ and  with $({\Lambda/ M_{Pl}})^4\sim 10^{-123}$ we get 
\be
\label{cepsc}
\epsilon\sim  10^{-120}.  
\ee
While this is a small number  the point to remember is that   $\epsilon$ is determined by the warp 
factor  at the bottom of the throat, eq.(\ref{defeps}) and this 
 is often exponentially  sensitive to fluxes, as for example happens in the Klebanov-Strassler \cite{Klebanov:2000hb} case.  
Thus modestly small ratios in flux can give the required
large hierarchy between the energy density in the quintessence field and the Planck scale and eq.(\ref{ccb})
can be met while taking the values of  $L,g_s$ to be  quite reasonable.  

In summary, we see that, as a first pass, 
 breaking the shift symmetry by  placing 5-branes in highly warped throats allows us to meet 
the requirements for a model of quintessence.   
\section{Other Terms in The Potential}

This is by no means the end of the story, however, for we have 
not  included the effects of supersymmetry breaking and 
moduli stabilization as yet. As was emphasized in the introduction these are expected to impose stringent constraints on any model. 
We will now turn to examining these constraints within the KKLT model for moduli stabilization 
\footnote{Another possibility is to use $\alpha'$ corrections for stabilizing moduli \cite{Balasubramanian:2005zx}.
We leave an investigation of our model  using such a mechanism for moduli stabilization for the future.
}.  
The conclusion will be that an additional term in the axion potential does indeed arise and typically dominates over the 
one we have kept above. However this term is also linear in the axion
 and by adjusting  the warp factor at the bottom of the two throats
our   model will be viable after all.

\subsection{Other contributions to the Axion Potential}

The first comment to make in studying the interplay of the moduli stabilization potential and the axion
is that the tree-level potential which arises in the presence of flux in IIB theory does not give rise to a potential for the $C_2$ and $B_2$ fields. This is in agreement with our general considerations about the shift symmetry not being broken at this order. 

There are two main possibilities to consider next. First, in the KKLT model
 non-perturbative effects are  used to stabilize
K\"ahler moduli. These effects might  give rise to a potential for the axion. 
Second, such a potential could arise due to the effects of the warp factor. The additional 
D3-brane charge that the axion gives rise to in turn back reacts on the geometry and 
can produce a change in the compactification volume. Since   the volume  modulus has been stabilized 
such a change  will come    at a cost in energy. 
We examine these two possibilities in turn below. 

Before doing so, let us ask in general terms when an additional  contribution to the axion potential,
$\delta V$ 
 can be neglected compared to the 
term we have already discussed, eq.(\ref{tension}).
Clearly the additional contribution should be small compared to the term we keep, 
\be
\label{condaneg}
\delta V \ll V. 
\ee
In addition the added term should not change the validity of the slow-roll conditions, eq.(\ref{slowroll}). 
If the corrections are polynomial in the axion, this is true automatically once eq.(\ref{condaneg})
is met and the slow roll conditions are imposed for the leading term. E.g., 
one of the slow-roll conditions gives 
\be
\label{exsr}
\left({\delta V'\over V}\right)^2 M_{Pl}^2 \sim \left({\delta V \over V}\right)^2 {M_{Pl}^2 \over \phi^2} \ll 1.
\ee
The slow-roll conditions due to  leading term requires that eq.(\ref{srlinear}) is valid. 
Eq. (\ref{exsr}) then follows from eq.(\ref{condaneg}). 
If the corrections are periodic in the axion, the same  slow-roll condition gives
\be
\label{exsrt}
\left({\delta V \over V}\right)^2 {M_{Pl}^2 \over f_a^2} \ll 1.
\ee
Using eq.(\ref{deffa}) we see that this condition is  more restrictive than eq.(\ref{exsr})
at weak coupling, when $g_s<1, L>1$. 
In the discussion below we will find that the corrections which can be neglected can be made sufficiently small so  
that their effects in  both eq.(\ref{condaneg}) and on the slow-roll conditions will  be small.

\subsection{Contributions From Moduli Stabilization}

\subsection{Corrections to the Superpotential}

As was mentioned above in a  KKLT model  type   situation the potential for the volume and other
 K\"ahler moduli arise from non-perturbative effects \cite{Kachru:2003aw}. Keeping only the overall volume this takes the form, 
\be
\label{npd3}
W_{np}=A e^{- \alpha T}.
\ee
Corrections to this superpotential which depend on the axion would generate an 
addition potential for it. 
We turn to examining them next.

The non-perturbative effects responsible for eq.(\ref{npd3})
 could arise from Euclidean D3-brane (ED3) instantons or they could 
arise due to gaugino condensation
in a $3+1$ dim.  non-Abelian gauge theory obtained 
 as the low-energy limit on a stack of  7-branes which wrap a 4-cycle
in the Calabi-Yau. 
For the ED3 case  it is possible that there are additional contributions
 which arise from bound states of 
Euclidean D1 branes and D3-branes. These do not seem to be suppressed   (at large volume) compared 
to the leading answer above \footnote{The ED1 charge can arise, for example,   
from induced world-volume flux along  a non-trivial two-cycle contained in the
 four-cycle wrapped by the ED3.  For fixed flux quanta the extra cost in energy for exciting 
this world volume flux vanishes as $L \rightarrow \infty$ suggesting that 
there is no extra suppression. }.
Including them  would therefore result in a superpotential of the form, 
\be
\label{rep2T}
W=B e^{-\alpha T} e^{c_1 G} 
\ee
where $a,c_1$ are coefficients.  The resulting contribution to the axion potential would then be 
\be
\label{deltapa}
\delta V \sim U_{\rm mod} \, a \sim M_{SB}^4 \,a
\ee
which is unacceptably large. 

To avoid this possibility it was suggested in \cite{McAllister:2008hb} that one consider the other possibility in KKLT models
and take    models where 
the non-perturbative corrections arise only due to gaugino condensation on wrapped 7-branes. 
This ensures that   additional contributions which can depend on the axion are highly suppressed. 

The suppression arises due to holomorphy and non-renormalization arguments involving RR axions. 
Let us recount the argument give in \cite{McAllister:2008hb} here for completeness. 
The non-perturbative superpotential  in the Yang-Mills theory is 
\be
\label{exsup}
W=A e^{-\alpha S}
\ee
where $S$ is the holomorphic gauge coupling. 
At large volume  
\be
\label{ymc}
S = T.
\ee
 Corrections to eq.(\ref{ymc}) which can induce a dependence
on the $C_2,B_2,$ axions  must vanish at large volume,
and therefore must be suppressed by inverse powers of ${\rm Re}(T)$. But by holomorphy this means they 
must also depend on the 
$C_4$ field which is the partner of the K\"ahler modulus. Now this $C_4$ field is also a type of an axion 
which arises 
 from the RR sector and  such a correction term would break the  shift symmetry of this field.
As we have argued above this cannot happen in perturbation theory. 
This means any correction to eq.(\ref{ymc}) must be exponentially suppressed in $T$ and this in turn would lead 
to a further exponential suppression  in the superpotential for the  axion dependent terms compared to the leading contribution in eq.(\ref{exsup}). 

The resulting  contribution to the quintessence  potential  from such a contribution would be 
\be
\label{delv2}
\delta V\sim U_{\rm mod} e^{-\beta\, {\rm Re}(T)} a.
\ee
Requiring that this correction is subdominant compared to eq.(\ref{v0}), gives, 
\be
\label{subdom}
U_{\rm mod} e^{-\beta\, {\rm Re}(T)} a < \Lambda^4
\ee
From eq.(\ref{umod}) we get 
\be
\label{reltwo}
e^{-\beta \, {\rm Re}(T)} a < \left({\Lambda \over M_{SB}}\right)^4 \le 10^{-60}
\ee
where we have used the fact that $M_{SB}\ge 1 \, {\rm TeV}$.
While the RHS is indeed small the exponential sensitivity of the LHS on the volume means this constraint can be easily met for the required range of the axion, $a \sim L^2/g_s$,  by taking modestly big values of $L$ and thus ${\rm Re}(T)$. 

\subsection{Corrections in the K\"ahler potential}

The  potential depends on both the K\"ahler potential and the superpotential so another source for 
additional 
terms in the axion potential  comes from corrections to the K\"ahler potential. 

 In fact  the K\"ahler potential in the tree-level theory   
itself involves mixing between the $T$ and $G$
modulus \cite{Grimm:2004uq} and as we see below this  typically gives rise to an unacceptably big 
contribution to the potential if the axion arises from a $B_2$ zero mode. This is why we are safer in using 
an axion coming from $C_2$, for which no such mixing arises at tree-level,  as the quintessence field.

The tree-level K\"ahler potential in the case we are considering  with one $T$ and $G$ modulus
is \cite{Grimm:2004uq}, 
\be
\label{kpmix}
K=-3 \log\left[T+\bar{T}+{3\over 2} e^{-\phi} C b b\right],
\ee
where $C$ is a coefficient determined by  the triple intersection numbers of four-cycles.  
 Note that only the $b$  field which arise from  $B_2$ 
 appears in the K\"ahler potential which is independent of the $a$ component coming from $C_2$. 
 Typically this mixing  will give rise to a quadratic term  for the $b$ axions,
\be
\label{massb}
\delta V \sim  U_{\rm mod}\, b^2. 
\ee
The slow-roll conditions requires $b>L^2$ so $\delta V \gtrsim M_{SB}^4$ and will therefore be much too big.
Since the mixing terms do not involve the $C_2$ axions they are safe in this respect. 

The K\"ahler potential will generically receive corrections beyond tree-level and these can break the 
$C_2$ shift symmetry.  
Since this  symmetry can only be
broken through quantum non-perturbative effects these effects must involve Euclidean D1-branes wrapping a
 two-cycle in the appropriate homology class. 
The resulting correction in the potential will be of the form 
given in eq.(\ref{delv2}) and will have a further exponential suppression in the volume. As discussed in 
eq.(\ref{reltwo})  such corrections
can be made small enough by taking  a modestly big value of ${\rm Re}(T)$.

 \subsection{Warping Effects}

Next we turn to warping effects. As we will see these will give rise to significant corrections.

The essential physics here is that the extra 3-brane charge which arises due to the presence of the 
axion gives rise to additional warping, and this warping in turn changes the 
 overall volume of the compactification
 \footnote{And more generally the size of all four-cycles.} \cite{Frey:2008xw} (for the 
warping corrections in K\"ahler potential, see also \cite{Giddings:2005ff,Shiu:2008ry,Douglas:2008jx}).
 Since the volume has been stabilized already 
this change comes at a cost in energy which depends on the warping and thus the axion.

A precise calculation capturing this effect  is difficult to carry out, at least with our current knowledge of flux compactifications. In the KKLT like scenario we are considering here, the potential for moduli stabilisation arises from non-perturbative effects. To obtain it one uses as input a K\"ahler potential and a 
superpotential, with  the non-perturbative effects being   incorporated in the superpotential.   
However once the 5 anti 5-brane system are included, SUSY is neccessarily broken and it is 
not so clear if the  resulting warping effects caused by the   axion can be included in this manner anymore. 
A first principlies calculation of the resulting changes in the potential is even more challenging. 
 
While a precise calculation seems hard,  a rough estimate which captures the essentially physics
above can be made as follows.   
Let the change in the internal volume, $V_I$, caused by warping due to the axion induced 3-brane charge 
  be $\delta V_I$.  Then the resulting potential for the axion can be estimated to be 
of order
\be
\label{delv3}
\delta V \sim U_{\rm mod}{\delta V_I \over V_I}
\ee
At first sight one might think that the change in the potential should be quadratic in the change in $\delta V_I$ and not linear in it, since the moduli have been stabilized and one is expanding around a minimum for 
them.  However,  a little more thought shows that this is not going to be typically true. 
 The reason is that the 
warping will not affect all terms in the potential in the same manner.
  As a result the potential itself changes once the effects of the 
warping are included.  

\subsubsection{A Subtlety}

Actually there is   one subtlety regarding the definition of the internal volume which we should address
before proceeding. 
For a  warped compactification of the type we are dealing with here with metric  
\be
\label{warp2}
ds^2=e^{2A(y)} dx_\mu dx^\mu+e^{-2A(y)} \tilde{g}_{ab}dy^ady^b,
\ee
the internal volume is  
\be
\label{intvola}
V'_I=\int d^6y \sqrt{\tilde{g}} e^{-6A(y)}.
\ee
However, the four dim. Planck scale, $M_{Pl}^2$, when expressed in terms of the string scale and $g_s$
is not proportional to $V'_I$. Rather, due to the warped nature of the geometry, we get 
\be
\label{vol}
M_{Pl}^2 \sim {1\over \alpha' g_s^2}  \int d^6y \sqrt{\tilde{g}} e^{-4 A(y)}.
\ee
This suggests that the natural variable for gravitational purposes,
 whose fractional change determines the change in the potential in eq.(\ref{delv3}) is 
 is really $V_I$
 defined by  
\be
\label{intvol2}
V_I \equiv   \int d^6y \sqrt{\tilde{g}} e^{-4 A(y)}
\ee
which appears on the RHS in eq.(\ref{vol}). 
Henceforth, this is the assumption we will make in 
 computing the corrections to the potential in eq.(\ref{delv3}) \footnote{Taking the fractional change in the volume  $V_I'$ instead does not change the central conclusions because it does not change the 
dependence on the warp factor $e^{A_0}$ for the resulting corrections.}. It is worth noting that  
 the warp dependent corrections to the K\"ahler potential calculated for SUSY situations in
 \cite{Frey:2008xw}  result in a correction are of the type in eq(\ref{delv3}), with this definition of $V_I$.

Now note that the internal metric in eq.(\ref{warp2}) is invariant under  a rescaling, 
$e^{2A}\rightarrow \lambda^2 e^{2A}$,
$\tilde{g}_{ab}\rightarrow \lambda^2 \tilde{g}_{ab}$. While this keeps the volume $V'_I$  unchanged
it changes  $V_I$. 
This is because  the warped metric eq.(\ref{warp2}) with this rescaling
 also changes lengths as measured in the non-compact directions,
and therefore  rescales $M_{Pl}$. 
Since we are only  interested in using $V_I$ to compute the fractional change
${\delta V_I \over V_I}$ in eq.(\ref{delv3}) though,
 any such ambiguity will cancel out in our calculation. 

In fact it will be safest for our purposes to fix this  rescaling ambiguity by setting  the volume of 
the unwarped metric to be unity in string units, 
\be
\label{volwarp}
\int d^6y \sqrt{\tilde{g}} =(\alpha')^3.
\ee
Then working with the resulting  expression for $e^{-4A}$, which is now well defined,
 we can  calculate $V_I$ and its change. 

To get going   first  consider  a situation where we ignore any effects of warping. 
$e^{-4A}$ is then a constant,
\be
\label{conse4A}
e^{-4A}=C_1.
\ee
The internal volume from eq.(\ref{intvola}), eq.(\ref{volwarp}), eq.(\ref{conse4A}) is 
\be
\label{intvolc}
V'_I=C_1^{3/2} (\alpha')^3.
\ee
Using our previous notation, eq.(\ref{intvol}), we get, 
\be
\label{prevn}
C_1^{3/2}=L^6. 
\ee
From the definition eq.(\ref{intvol2}) we then  get that  
\be
\label{valvol}
V_I=C_1=L^4 (\alpha')^3.
\ee

\subsubsection{The Leading Effect}

We are now ready to estimate the change in $V_I$ due to warping effects. 
Consider a simple model first of a stack of $N$ D3-branes   at some location in the 
internal space.  For simplicity we take the internal geometry to be flat, i.e., a torus.
The geometry is of the form,   
\be
\label{geomv}
ds^2=e^{2A} dx_\mu dx^\mu+ e^{-2A} (dr^2 + r^2 d\Omega_5^2).
\ee
The change in the warp factor produced by the D3-branes is,
\be
\label{wfv}
\delta e^{-4A}=  {R^4\over r^4}, 
\ee
where $r$ is a radial coordinate in the internal space measuring distance from the branes and $R$ is the $AdS_5$ radius,
\be
\label{rad}
R^4= 4\pi g_s N \alpha'^2.
\ee
Since we have set the unwarped volume to be unity in string units, eq.(\ref{volwarp}), the 
radial variable $r$ in eq.(\ref{geomv}) has a range $0\le r \lesssim O(\sqrt{\alpha'})$. 
The correction to the warp factor are well described by eq.(\ref{wfv}) for  $r\ll O(\sqrt{\alpha'})$,
when the effects of the ``image '' stacks needed to implement the boundary conditions on the 
torus are not important.  Near the stack of branes, as $r\rightarrow 0$,
 the constant term is relatively   unimportant 
and $e^{-4A}$ is given by eq.(\ref{wfv}) resulting in $AdS_5\times {\bf S}^5$ space.

The change in the internal volume caused by the warp factor can now be calculated as
\be
\label{chnginta}
\delta V_I=\int d^6y \sqrt{\tilde{g}} \delta e^{-4A}
\ee
Using, eq.(\ref{wfv}) gives, 
\be
\label{cbrane}
\delta V_I \sim R^4 \alpha'   \sim 4 \pi g_s N L^2 \alpha'^3.
\ee

Now actually in our model  what we are interested  in is the
change in volume caused by the axion.  The D3 charge that the axion induces
is 
\be
\label{chnga}
\Delta N \propto a.
\ee
From eq.(\ref{valvol}) and eq.(\ref{cbrane}) we see that the fractional 
change in $V_I$ caused by the axion is then, 
\be
\label{axdep}
{\delta V_I \over V_I}  \sim  g_s {a \over L^4}.
\ee

The resulting contribution to the potential this gives is 
\be
\label{chnpota}
\delta V\sim U_{\rm mod} \ g_s \  {a\over L^4}.
\ee
It is easy to see that this is unacceptably large. 
Requiring that 
\be
\label{relccd}
\delta V < \Lambda^4
\ee
gives the condition
\be
\label{condrelcc}
U_{\rm mod} g_s  {a\over L^4} <\Lambda^4.
\ee
Let us set $g_s \sim O(1)$, with 
$U_{\rm mod}\sim M_{SB}^4$, and $a \sim L^2$, eq.(\ref{condrelcc})   then gives, 
\be
\label{reldd}
L>  {M_{SB}^2 \over \Lambda^2} \sim 10^{30} \left({M_{SB}\over 1 \, {\rm TeV}}\right)^2. 
\ee
Now $L=10^{30}$ leads to a huge internal space and it is easy to see that the resulting string scale is ridiculously low. 

\subsubsection{Incorporating The $\overline{\uD 3}$-branes}

Our analysis is in fact incomplete for one obvious reason. We have so far only included the effect due to the NS5-brane
where the axion induces  D3-brane charge and tension. In the actual situation at hand there is also the anti 5-brane 
where $\overline{\uD 3}$-brane charge and tension  is induced.  
Including its effect   can  cancel the contribution found above to leading order, but not exactly, as we will see. 

It is best to work in a situation where the ambient $N$ units of 
 3-brane charge  in the throats in which the 
5-branes are placed is much bigger than the 3-brane/anti 3-brane charge induced by the axion.
One can then estimate the effects of the axion to leading order in $a/N$. 
The presence of the ambient 3-brane charge and related five-form flux breaks the symmetry between the backreaction effects
of the induced 3-brane charge   on the 5-brane and the anti 3-brane charge induced on the anti 5-brane,
as we will see.

It is useful to estimate the contributions of the $\overline{\uD 3}$-brane charge and tension in two steps. 
The effects of the $\overline{\uD 3}$-branes arise from localized sources in the metric and five-form equations. 
The charge of an  $\overline{\uD 3}$-brane is opposite to that of a D3-brane while its   tension is the same. 
It is helpful in making our estimates to think of the $\Delta N$ $\overline{\uD 3}$-branes as a sum of two kinds of sources \cite{Maldacena:2001pb}. 
The first kind is $\Delta N$ 3-branes with both charge  and tension opposite to that of a D3-brane. In terms of sources
these can be thought of as $-\Delta N$ D3-branes. The second kind of source
 are $\Delta N $ pairs of D3 and $\overline{\uD 3}$-branes, with  each pair  together having no net charge and twice the tension of
 a D3-brane. We will see that the leading effect arises due to the first kind of source and this is exactly canceled,
 in  a symmetric situation, by the contribution coming from the D3-branes in the other throat.  
 
This leaves the contribution from the $\Delta N $ pairs of 3 and anti 3-branes. The system of D3-brane $\overline{\uD 3}$-brane pairs 
placed at the bottom of a KS throat was studied in \cite{DeWolfe:2008zy,McGuirk:2009xx,Bena:2009xk},  see also \cite{Kachru:2002gs,Kachru:2003sx,Bena:2010ze}.  One important change is that the geometry is no longer of the type eq.(\ref{warp2}) 
with $\tilde{g}_{ab}$ being the metric of
the Calabi-Yau manifold. Rather the backreaction in this case distorts the metric by more than an overall warp factor.
The ambient five-form flux, it was found,  screens the  effects of these pairs at large distance. 
As a result their resulting contribution is suppressed and subdominant.  

Below we first discuss the contribution due to the $-\Delta N$ D3-branes, and then turn to the effect of the pairs later. 

\begin{figure}
  \begin{center}
  \includegraphics[width=5cm]{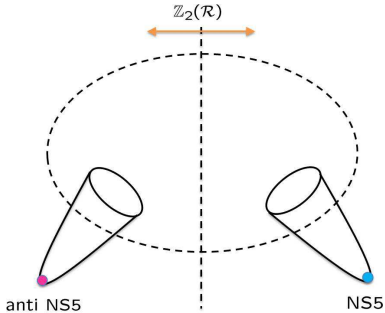}
  \end{center}
  \caption{\footnotesize A schematic diagram  of the internal space showing the 
two throats related by the ${\mathbb Z}_2$ ${\cal R}$ symmetry containing the NS5-brane and the $\overline{\rm NS 5}$-brane respectively.}
  \label{fig:orbifolding}
\end{figure}

Including both the $\Delta N $ D3-branes located at $x_0$
 and the $-\Delta N$ D3-branes, which originate from the $\overline{\uD 3}$-branes
 and are taken to be at $\tilde{x}_0$, leads to the equation for the warp factor,
\be
\label{weq}
\tilde{\nabla}^2  \delta e^{-4A}= C_1 \Delta N \left({\delta^6(x_i-x_{0}) \over \sqrt{\tilde{g}(x_0)}} 
- {\delta^6(x_i+x_{0}) \over \sqrt{\tilde{g}(\tilde{x}_0)}}\right).
\ee
The opposite relative sign  mean that they contribute oppositely to the change in the  $V_I$,  eq.(\ref{intvol2}).
In general the contributions will not exactly cancel, the residual contribution would then typically still be of order 
eq.(\ref{chnpota}) and unacceptably big. However if there is a ${\mathbb Z}_2$ symmetry ${\cal R}$ as in figure \ref{fig:orbifolding}, of the
 Calabi-Yau space,
 under which the two throats are exchanged,   which 
is also respected by the ambient flux  and  if this symmetry
is then  only broken by the brane-anti brane pair,  then it is easy to see that 
  the two contributions will exactly cancel. This was the reason for 
assuming such a symmetry when we described the basic setup in\footnote{Ref. \cite{Flauger:2009ab}
 also discusses the use of a   ${\mathbb Z}_2$ symmetry for similar purposes.} \footnote{Another
possibility, not shown in Figure 1, is that  the two throats are themselves contained in a warped region - the 
``parent'' throat.
In the IR the parent throat  splits into the two 5-brane containing throats.}   \S3.

\subsubsection{The Subleading Effect Due to pairs}

The subleading effect is due to the $\Delta N $ brane-anti brane pairs. The resulting change in the geometry 
was  calculated in \cite{DeWolfe:2008zy} for UV of the KS throat, and \cite{McGuirk:2009xx,Bena:2009xk} for IR. As was mentioned it is not simply of the warped form, eq.(\ref{warp2}).
Nevertheless an estimate of the resulting change in volume can be obtained by just keeping the change in the warp factor.
 
In the region far from the tip of the KS geometry where the pairs are located, the geometry is essentially $AdS_5\times {\bf S}^5$ and the perturbation  goes like, 
\be
\label{deltavthree}
\delta e^{-4A} \sim \Delta N e^{4 A_0} R^4 {g_s \alpha'^2 \over |r|^8}
\ee
where the $e^{4A_0}$  factor arises from the warping at the tip of the throat.
 Note the $1/|r|^8$ fall off which is much faster than the 
$1/|r|^4$ fall-off for a D3-brane source \footnote{Actually there is a  $\log(r)$ factor on the  RHS but we neglect
this below. This factor arises because the total D3 charge increases in the KS solution as one
goes to large $r$. For our crude estimate we can neglect this effect. Instead we model the total flux in the throat to be roughly constant and of order say $N$. $R$ which appears on the RHS then is 
related to this flux by eq.(\ref{rad}).}. The important point here is that the brane-anti brane pairs
give rise to a normalizable deformation in the asymptotic $AdS_5\times {\bf S}^5$ region, which falls like
$1/|r|^8$, this deformation is suppressed by $e^{4A_0}$ because the warp factor at the bottom of the throat suppresses the energy density of the pair in this manner. 
While some of the discussion in \cite{DeWolfe:2008zy} is in the context of the KS geometry their
main result is more general and should apply to other  situations as well 
which are asymptotically $AdS_5\times {\bf S}^5$, with   the warped throat terminating in a  region
where the warp factor attains a minimum value $e^{A_0}$.

The resulting change in $V_I$  that  eq.(\ref{deltavthree})  leads to can be calculated from
 eq.(\ref{intvol2}). 
Since our  purpose is to make a rough estimate we may as well approximate the geometry to be a region of $AdS_5\times {\bf S}^5$
with  the $AdS_5$ being cut-off both in the UV and the IR;  i.e., of the type considered in the RS1 model, with metric, 
\be
\label{adspart}
ds^2={r^2\over R^2} (dx_\mu dx^\mu) + { R^2 \over r^2} (dr^2 + r^2 d\Omega_5^2)
\ee
where the radial coordinate has range $r_{\rm min} \le r \le r_{\rm max}$. This metric has the form, 
eq.(\ref{warp2}), with
\be
\label{metflat}
\tilde{g}_{ab} = \delta_{ab}
\ee 
being
the flat-space metric.  
The warp factor at the bottom of the cut-off $AdS$ region is ${r_{\rm min}^2 / R^2}$ which in this crude model 
should be equated with 
$e^{2A_0}$ at bottom of the KS-like throat giving,
\be
\label{thb}
r_{\rm min}^2=R^2 e^{2A_0}.
\ee

The brane anti brane pair located at the bottom of the throat produces a  change in the warp factor
 eq.(\ref{deltavthree})
and we interested in the change in $V_I$ this leads to.  Integrating eq.(\ref{intvol2}) using eq.(\ref{metflat})
gives, 
\be
\label{inta}
\delta V_I\sim \Delta N e^{4A_0} R^4 \alpha'^2 {g_s \over r_{\rm min}^2},
\ee
where we have taken $r_{\rm max} \rightarrow \infty$ since the integral converges in the UV. 
Next using eq.(\ref{thb}) gives, 
\be
\label{delfive}
\delta V_I \sim \Delta N e^{2 A_0} R^2 \alpha'^2.
\ee

So far we have only estimated the contribution from the region far from the tip of the throat where the 
pairs of 3-branes are  located. 
We can also try to estimate  the contribution from the region near the tip as follows.  Let us continue to crudely
model the throat as a cut-off $AdS_5\times {\bf S}^5$ geometry, eq.(\ref{adspart}) with the  pairs located  at $r_{\rm min}$.
Very close to the sources, it seems reasonable to estimate that the change produced in the warp factor
by a D3-$\overline{\uD 3}$ pair is of order  the change produced by a single $\overline{\uD 3}$-brane.
 This results in a change in the warp factor,
\be
\label{paradel}
\delta e^{-4 A} \sim \delta N {g_s \alpha'^2 \over r^4}  
\ee
and a change in $V_I$ 
\be
\label{paraeight}
\delta V_I \sim \delta N {g_s \alpha'^2} R^2 e^{2A_0},
\ee
where in obtaining eq.(\ref{paraeight}) this time, since we are interested in the 
region close to the brane, we cut off the integral before
$r$ gets too large at $r\sim r_{\rm min}$. 
Note, that this  result is 
 parametrically of the same form as the contribution from the region far from the tip
 eq.(\ref{delfive}).  Adding the effects of the near-and far regions we would therefore expect an answer
of the form eq.(\ref{delfive}).


To complete our analysis  we now calculate the resulting term in the axion potential. 
Since $\Delta N \propto a$ we get from eq.(\ref{delfive}), eq.(\ref{valvol}),eq.(\ref{delv3})  that the extra potential generated for the axion is
\be
\label{extrapot}
\delta V \sim U_{\rm mod} e^{2 A_0} {R^2 \over\alpha'  L^4} a.
\ee

The estimate for the near-region can be improved in the context of a KS throat using the results of 
\cite{McGuirk:2009xx,Bena:2009xk}. This more careful analysis gives results which essentially agree with eq.(\ref{paraeight}), eq.(\ref{extrapot}).

In the above discussion we have not kept track of  signs and coefficients. 
   The contribution due to warping   eq.(\ref{delfive}), 
we will see below, will
typically lead to a contribution in the axion potential that dominates over other contributions, 
including the term due to the induced tension of the D3, $\overline{\uD 3}$-branes we first considered in eq.(\ref{v0}). 
It will be important for the axion model to work  that this net contribution is positive. Our analysis above is too preliminary to allow us to determine this sign. 
Physically one would expect the net contribution to be positive, since otherwise the added D3-brane
charge on the 5-brane anti 5-brane would reduce the energy of the system and this energy would decrease as the charge increased. However, one should clearly try to determine this  from a first principles calculation, along 
with improving the other aspects  of this calculation as well. We  leave this  for  future investigation. 
\subsubsection{Final Conclusions}
The changes in $V_I$, eq.(\ref{intvol2}), that arise due to  warping  could  possibly manifest themselves
  in corrections to 
 the superpotential and/or  the K\"ahler potential. 
For example,  ${\rm Re} (T) \sim L^4$ in the unwarped case and it is more generally  related to the size of the non-trivial 
four-cycle. Thus $T$  could  change once warping effects are included causing in turn changes in the superpotential and 
the K\"ahler potential \footnote{We thanks L. McAllister for a discussion on this issue.}. 
The warping induced changes might also lead to effects that cannot be incorporated into 
supersymmetric data in such a ready fashion, since the induced $\overline{\uD 3}$-brane charge breaks SUSY. Either way the resulting 
corrections to the axion potential will be given  by eq.(\ref{extrapot}), which is the main result from this
 subsection. 

\subsection{ An Additional Contribution}
Before concluding this section let us  discuss one additional contribution which will turn out to be 
unimportant.

There is a correction to the axion potential which arises due to the interaction 
between the 5-brane and the anti brane in the two different throats. In the limit where the induced 
D3-brane tension dominates over the 5-brane tension this is simple to estimate. The potential reduces to that calculated between D3 and $\overline{\uD 3}$-branes calculated in \cite{Kachru:2003sx}. This gives a result,
\be
\label{intreaction}
\delta V \sim {a^2 \over |r_1-r_2|^4} \epsilon^2,
\ee
where the two throats are located at radial coordinates $r_1,r_2$ and for simplicity we have taken the
 internal space to be flat. Here,  $\epsilon = e^{4A_0}$, is the warp factor at the bottom of the throat. 
We see that this contribution is suppressed by an extra factor of $\epsilon$ compared to the tension
term calculated in eq.(\ref{tension}).
Since $\epsilon$ is very small,  as the preliminary analysis which actually underestimates $\epsilon$ already found in 
eq.(\ref{cepsc}), we see that this contribution will be highly suppressed compared to the ones we keep. 

\section{The Model: A More Complete Look}
To summaries, we started with a potential of the form, eq.(\ref{v0}),
\be
\label{pot0}
V_0=c_1 {1\over \alpha'^2 g_s} e^{4A_0} a.
\ee
Then we  found that corrections will generate an additional term, eq.(\ref{extrapot}). Using the relation eq.(\ref{umod}) this can be written 
in terms of the SUSY breaking scale $M_{SB}$ as,
\be
\label{pot1}
V_1=c_2 M_{SB}^4 e^{2A_0} \left({R^2\over \alpha' L^4}\right) a.
\ee
Here $L^6$ is the volume in string units, and $R$  is roughly speaking the radius of the AdS-like throat regions in which the 
5-brane and anti brane's are placed \footnote{ More correctly if the throats are of  KS type \cite{Klebanov:2000hb}  the five-form flux changes 
along them and $R$ can be taken to be   an appropriate average for this radius.}. In the supergravity approximation
\be
\label{relra}
R^2/\alpha' >1.
\ee
We also note that   $e^{A_0}$ in eq.(\ref{pot0}), eq.(\ref{pot1}), is the warp factor at the bottom of the throat.

The total potential is then linear in the  axion  and the sum of the two terms, 
\be
\label{totpot}
V=V_1+V_1=\mu^4 a
\ee
where the final expression on the RHS can be taken to be  the definition of the scale $\mu^4$. 

We should also note that eq.(\ref{pot0}) and eq.(\ref{pot1}) are valid only for $a>0$ and also
for  sufficiently large 
values of $a$. The more correct form for $V_0$ is in eq.(\ref{tension}), we will return to 
a discussion of corrections to the linear form of the potential in \S6.3.

\subsection{The Energy  Scales}

To understand which of the two terms, eq.(\ref{pot0}) or eq.(\ref{pot1}) is bigger let us start by first setting
$g_s=L=1$.  This also sets $1/\alpha'^2 \sim M_{Pl}^2$. 
Also we set ${R^2\over \alpha'}=1$.  Since we have not been able to calculate $c_2$ anyways, we  also   
 neglect any dependence on the coefficients $c_1,c_2$. Finally we set $a \sim O(1)$ so that $a \sim L^2$ as is needed for the slow-roll conditions to be met eq.(\ref{rangeaxion}).    
Setting the moduli stabilization contribution $V_1$  to be of order the vacuum energy density, we then get,
\be
\label{conmv}
V_1 \sim M_{SB}^4 e^{2A_0} \sim \Lambda^4.
\ee
This gives $V_0$ to be  
\be
\label{estv0}
V_0\sim {M_{Pl}^4 \Lambda^4 \over M_{SB}^8} \Lambda^4.
\ee

Now if the SUSY breaking scale is 
$M_{SB}\sim  1 \,{\rm TeV}$,
\be
\label{relaa}
{M_{Pl}^4 \Lambda^4 \over M_{SB}^8}\sim O(1).
\ee
 Thus 
$V_0\sim V_1$ and both make roughly an equal contribution to the vacuum energy. 
However, as the moduli stabilization scale and SUSY breaking scale are   raised we see that $V_0$ begins to become smaller than $V_1$ and less important. 
For example, when $M_{SB} \sim 10^{10} \, {\rm GeV}$, the intermediate scale, $V_0$ is much smaller,  
\be
\label{newv0}
(V_0)^{1/4}\sim 10^{-17} \, {\rm eV}.
\ee
 
Let us keep track of  the dependence on the volume $L$ and $g_s$ also now. We set $a\sim L^2/g_s$, eq.(\ref{srlinear2}). 
Setting $V_1$ to be of order the energy density gives,
\be
\label{estv1p}
V_1\sim  M_{SB}^4 ({R^2 \alpha'}) e^{2 A_0} {a\over  L^4} \sim M_{SB}^4 ({R^2 \alpha'}) e^{2 A_0} 
{ 1\over L^2 g_s} \sim \Lambda^4.
\ee
Solving for $e^{2A_0}$ gives, 
\be
\label{vale2a}
e^{2A_0}\sim \left({\Lambda^4 \over M_{SB}^4}\right) g_s L^2 \left({\alpha'\over R^2}\right).
\ee
Using the fact that $M_{Pl}^2 \sim {L^6 / (\alpha g_s^2)}$ and substituting in eq.(\ref{pot0}) then gives, 
\be
\label{estv0p}
V_0\sim  {g_s^4 \over L^6} \left({\alpha'\over R^2}\right)^2 \left({M_{Pl}^4 \Lambda^4 \over M_{SB}^8}\right) \Lambda^4.
\ee
For our approximation of classical supergravity to be valid, $g_s<1, L>1,$ and $R^2>\alpha'$. This means even for 
$U_{\rm mod}\sim ({\rm TeV})^4$,  $V_0<\Lambda^4$ and thus $V_0$ will be subdominant. 
Increasing $L, R^2,$ or decreasing $g_s$ will make $V_0$ even smaller. Similarly increasing $U_{\rm mod}$ will 
also reduce $V_0$. 

Our conclusion is that the second term $V_1$, which arises due to the interplay of moduli 
stabilization and warping caused by the axion getting turned on, is typically the dominant contribution. 
 We must caution the reader that we have not kept track of  numerical coefficients, 
some of these could be large, and our statements here are really only  parametric nature. 

The warped down string scale at the bottom of the throats where the 5-branes are present is,
\be
\label{wdsscale}
m_s \sim  {1\over \sqrt{\alpha'}}  e^{A_0} \sim {M_{Pl} g_s \over L^3} e^{A_0},
\ee
where in obtaining the last relation we have used eq.(\ref{mpl}). 
Supergravity modes (KK modes in the throat region)  have a mass which is even lower by a factor of $\sqrt{\alpha'}/ R_0$
where now $R_0$ is the  radius of curvature at the bottom of the throat \footnote{This might  well  
be smaller than $R$ as defined above.}. 

For the case we discussed first above, around eq.(\ref{estv0}), with $g_s\sim L\sim  a \sim 
R_0\sqrt{\alpha'} \sim O(1)$ and $M_{SB}=1 \, {\rm TeV}$
we see that the warped down string modes and KK modes have a mass of order $10^{-3} \, {\rm eV}$. As $M_{SB}$
rises the mass becomes even lower. For $g_s\sim L \sim a \sim R_0\sqrt{\alpha'} \sim O(1)$ 
and $M_{SB}\sim 10^{10}\, {\rm GeV}$ this mass is of order 
$10^{-17}\, {\rm eV}$. Thus there are many very light particles which arise due to string modes and KK modes in the highly
 warped region. We will have more to say about these light particles in \S6.1.

\subsection{The Constraints}
It is also worth summarizing the constraints on the various parameters of the model. 

The slow-roll conditions require that the axion satisfy the condition, eq.(\ref{rangeaxion}).
Our estimate  for the $V_1$ term in the potential is valid only when the charge contributed by the axion is a small 
fraction of the total charge supporting each throat, this gives the condition, 
\be
\label{condchrg}
a\ll N,
\ee
Using  eq.(\ref{rangeaxion}) and eq.(\ref{rad}) this give,
\be
\label{condl}
L^2 \ll R^4/\alpha'^2.
\ee
Finally the total volume of the internal space is $L^6 (\alpha')^3$, this must be bigger than the volume of
 each warped throat $\sim R^6$,
leading to 
\be
\label{condlt}
L^4\gg R^4/\alpha'^2.
\ee

The summary is that the  conditions,
\be
\label{condfinal}
L^2<g_s a \ll R^4/\alpha'^2 \ll L^4
\ee 
 must  all be  met. They are mutually compatible so there is no obstruction to doing so.  
In addition the scale $\mu^4$  defined in eq.(\ref{totpot}) must meet the condition 
\be
\label{condmuf}
\mu^4 a =\Lambda^4
\ee
where $\Lambda$ is defined in eq.(\ref{deflambda}). 
This does require $\mu$ to be a very small energy, but as was emphasized in \S3.2, where a preliminary
 estimate for $\mu$ was carried out by neglecting $V_1$, this can be arranged by choosing a modestly small ratio of fluxes
which results in an exponentially small value of $e^{A_0}$. 

\subsection{Additional Comments}

Let us end  this section with two comments.  

First, it could be that despite our best attempts at being careful we have missed  some important
 additional contributions to the axion potential.  
The following general reasoning suggests that even if this were true, incorporating such effects would
probably lead to a workable model within the kind of setup we have explored. 
One would expect that  any additional contribution is  linear in $a/N$,
as long as $a\ll N$.
 Also, since the shift symmetry is broken by the 5-branes
whose effects are suppressed by $e^{A_0}$, such a term should also be suppressed by a positive power of $e^{A_0}$. If this power is smaller than $2$, then this additional contribution could dominate over
the terms we have kept. However in this case we can simply adjust the warp factor so that the resulting
 scale $\mu$ in  eq.(\ref{totpot}),
 after accounting for this term, has the correct value. This will typically make  
$e^{A_0}$ even smaller and thus the particles as the bottom of the warped throats even lighter,
but, as we will see in \S6.1, this is an acceptable price to pay since these particles do not 
lead to any observable signals anyways.  

Second, 
 in the KKLT construction the negative vacuum energy 
density that arises after moduli stabilization
 is canceled by the addition of $\overline{\uD 3}$-branes which are placed in a warped throat. 
It is natural to ask whether in our model one can dispense with these $\overline{\uD 3}$-branes and their 
attendant throat, 
and instead cancel the negative vacuum energy by the axion  dependent contributions, eq.(\ref{totpot}). 

A simple calculation shows though that this would require too many units of flux $N$ stabilizing the 
5-brane throats. 
Let the negative energy generated by moduli stabilization  be $-C$,
including it in  the total potential gives,
\be
\label{inctp}
V=\mu^4a -C.
\ee
The constant  $C$  can be absorbed by shifting the axion 
\be
\label{shiftax}
a \rightarrow a + \Delta a
\ee
with \be
\label{abs}
\Delta a= C/\mu^4.
\ee
The resulting analysis 
in terms of the shifted axion  then  exactly agrees with  what we had done earlier, when $C=0$. 
Since the shifted axion must satisfy the condition,  eq.(\ref{rangeaxion}), 
(to get a model of quintessence)
 equating $V$ with 
$\Lambda^4$ now will again lead to the conclusion that 
\be
\label{conmu}
\mu^4<\Lambda^4,
\ee
for $L>1, g_s<1$. 
The resulting shift is therefore
\be
\label{shifttwo}
\Delta a ={U_{\rm mod}\over \mu^4}>{({\rm TeV})^4 \over \Lambda^4} =10^{60},
\ee
where we have used the fact that $C=U_{\rm mod}\ge O({\rm TeV})^4$.
The linear potential we have used though is valid only when  eq.(\ref{condchrg})
 is true and   this condition involves  the total value of the axion including its shift.
 Thus the large value of $\Delta a$ would require a large  amount of flux $N$, 
which makes this idea  implausible. 

Our model therefore must incorporate the two throats with the 5-branes and at least one 
additional throat (or perhaps two which are related by the ${\mathbb Z}_2$ symmetry) to contain the $\overline{\uD 3}$-branes of the KKLT setup.

\section{Some Features of The Model}
We now turn to  discussing some general features of the model. 

\subsection{Many Light Particles:}
The model has many light particles. We saw that the warped down string scale at the bottom  of the two throats where the 5-branes
are located is at least $10^{-3} \, {\rm eV}$ and typically is much lighter for a   scale of SUSY breaking 
$M_{SB}>1 \, {\rm TeV}$ . Thus there are many very light particles in the theory which come from string modes and   Kaluza Klein modes
localized in the warped region. 
 
One might be worried that so  many light particles would be in conflict with observation. However, this is not so. 
The model is in fact closely related to the Randall-Sundrum II model (RS II) \cite{Randall:1999vf}
 in which there is only one brane called the Planck brane. 
More accurately it is akin to a Randall Sundrum I model (RS I) \cite{Randall:1999ee} but with the warped down energy scale at the ``Standard Model'' brane
being $10^{-3} \, {\rm eV} $ or much lower.  The RS II model can be thought of as a limit of RS I where the standard model brane is moved away to infinity resulting in a non-compact dimension. 

We take the standard model degrees of freedom to not live in the two throat regions where the 5-branes are located. 
In terms of the RS model they are then degrees of freedom on the Planck brane. The four-dim. graviton is a non-normalizable mode in $AdS_5$ and is localized on the Planck brane as well. The first thing to note is that the throat regions make a finite 
contribution, of order $R^6$, to the volume of the internal space where $R$ is given in eq.(\ref{rad}) and $N$ are 
the number of five-form flux units supporting the throat. As a result the $4$ dim. Newton's constant is finite and as we have discussed above can take the required value consistent with the other constraints of the model. In addition the couplings of the matter fields with the KK modes in the throat are suppressed so that corrections to Newton's gravitational law due to KK exchange
and corrections to energy loss in  gravitational radiation are suppressed by a factor of
 $(p R)^2$ compared to the leading answers which neglect these KK modes, where $p$ is the characteristic energy scale involved.
Now in the model   the flux units $N$ and thus $R$ has to be big enough to meet the condition eq.(\ref{condfinal}),
 however this can be typically arranged by taking $R$ to be not very different from the Planck scale. As a result this suppression is very significant
and results in no detectable signal.  E.g., taking $N\sim 1000$, $g_s \sim O(1)$, eq.(\ref{condfinal}) can be met for 
$L \sim O(10)$, and $a \sim O(100)$. In this case, $R^{-1} \sim 10^{16} \, {\rm GeV}$ so that the suppression is of order $(p/10^{16}\, {\rm GeV})^2$
and is indeed very significant.  Finally as argued in \cite{Randall:1999vf} the  interactions between the four-dim. graviton and KK modes
are also highly suppressed, because the four-dim. graviton is localized on the Planck brane and a typical KK mode
is localized deep inside the $AdS$ region.

Besides light particles in the two throats,  there are also  moduli, not localized in the throats,   
which are stabilized by $U_{\rm mod}$ (e.g., the overall volume). These   have a mass
\be
\label{maddmod}
m_{\rm moduli}^2\sim {U_{\rm mod} \over M_{Pl}^2}. 
\ee
For $U_{\rm mod}\sim {\rm TeV}^4 $ this is also a mass of order $10^{-3}$ \, {\rm eV}. As the scale of moduli stabilization is increased the
masses of these moduli increase, while, as we have seen above, typically 
 the masses of the warped down string states and the KK states move down. 

In constructing a complete cosmological model one would have to deal with the moduli problem 
\cite{Banks:1993en,de_Carlos:1993jw} that arises due to all 
these light modes. Perhaps after high-scale inflation one can arrange so that 
the KK and string  modes in the warped throats are not significantly excited and 
the moduli like the overall volume  which are spread out over the whole CY  are either heavy enough to not 
cause a problem, or  are  rendered unproblematic by symmetry considerations, \cite{Dine:1995uk}, or by thermal 
inflation \cite{Lyth:1995ka}. 

In particular it is important to ensure that the two throats are not excited to even small finite temperature. This would cause a black brane metric to replace the throat geometry at the tip, the two NS5-branes would then fall into the horizon of the black brane and the resulting
equation of state would be more akin to a thermal gas and unsuitable for quintessence. 
This is especially a concern because of the low warped down energy scale at the bottom of the throat. 

If reheating after inflation results in entropy  being dumped 
into standard model fields, this can probably be ensured.
The suppressed interactions with the throat excitations will then prevent the throat degrees of 
freedom from coming into equilibrium with the standard model degrees as long as the temperature 
after reheating is somewhat lower than the GUT scale, and this can prevent the formation of a black 
brane horizon \footnote{We thank S. Kachru for discussion in this regard.}.

\subsection{Rotation of The CMB Polarization}
It is well known that an axion $a$, which is a  pseudoscalar, can couple to the photon through the 
\be
\label{dell}
\Delta L = \zeta a F\tilde{F}
\ee
term in the 
Lagrangian causing the E mode of the CMB to be rotated into the B mode \cite{Carroll:1989vb,Lue:1998mq,Arvanitaki:2009fg}. Here $\zeta$ is a constant coefficient.
Such a coupling also  causes the direction of linearly polarized
light to rotate in the course of propagation from a  distant source \cite{Harari:1992ea}. 

The extent of rotation of the E mode to the B mode is parameterized by an angle 
\be
\label{angle}
\Delta \alpha \sim \zeta \Delta a
\ee
where $\Delta a$ is the total change in the axion due to its time evolution. 

CMB experiments put a  bound  on $\Delta \alpha$. The bound in \cite{Komatsu:2010fb} is  
\be
\label{boundrot}
\Delta \alpha = -1.1\deg \pm 1.3\deg ({\rm stat.}) \pm 1.5\deg ({\rm syst.}) \ \ \  (68\%\  {\rm CL})
\ee
which is  currently consistent with $\Delta \alpha$ vanishing. 

A significant limitation of  this paper is that we have not   discusses  how the standard model arises
in our construction.  As a result we do not know  whether a coupling of the type in 
 eq.(\ref{dell}) in fact arises\footnote{Such a coupling between the axion of interest and $SU(3)$ color gauge field must be absent otherwise the resulting QCD induced 
potential for the axion would be  much too big.}. 
In fact it is more or less clear that such a coupling cannot arise in a SUSY preserving manner. For that to happen the SUSY partner
of the axion would have to play the role of the QED coupling constant. Typically this happens only if the axion partner 
 is a non-compact scalar.  In our model where  the axion arises from $C_2$, however, 
 its  partner arises  from $B_2$ and therefore is a compact scalar. 

The coupling to electromagnetism could arise though in a SUSY non-invariant way and this could well  be allowed if 
 SUSY is broken at a sufficiently high scale. 
In particular, such a coupling would arise if the $U(1)$ gauge group of the SM  arose from a 
D5-brane wrapping the same two cycle \footnote{Or rather  the same  appropriate orientifold even combination of two-cycles.} which 
gives rise to the zero mode for the  axion. On the world-volume of the D5-brane would be the couplings,
\be
\label{ld5}
S=(2\pi \alpha'^2) T_5 \left[\int d^6x \sqrt{-g}{1\over 4} F_{\mu\nu}F^{\mu\nu}  + {1\over 2 } \int C_2\wedge F \wedge F\right].
\ee
Reducing to $4$ dimensions gives, 
\be
\label{fourdimred}
S=(2\pi \alpha'^3) T_5\left[ L^2 \int d^6x \sqrt{-g} {1\over 4} F_{\mu\nu}F^{\mu\nu} + {1\over 2} a \int F \wedge F\right].
\ee

Now redefining $F_{\mu\nu}$ so that the gauge kinetic term is canonical we see that the coupling $\zeta$ in eq.(\ref{dell})
 is 
\be
\label{valzeta}
\zeta={1\over L^2}.
\ee
As a result 
\be
\label{delalpha}
\Delta \alpha \sim {1\over L^2} \delta a.
\ee

Using the observed limits in eq.(\ref{boundrot}) (converted to radians)  we  then get 
\be
\label{cenval}
{\Delta a \over L^2} <10^{-2}.
\ee
The canonically normalized field is $\phi=f_a a \sim {a \, g_s M_{Pl}}/L^2$ as in eq.(\ref{defaa}). The change in $\phi$ is 
given in eq.(\ref{changea}), using  eq.(\ref{cenval}) this  gives, 
\be
\label{cenval2}
{M_{Pl}\over \phi} \lesssim 10^{-2} g_s \lesssim 10^{-2},
\ee
where we have used the fact that $g_s <1$. 
This tells us that to be in agreement with the bound eq.(\ref{boundrot}) the canonically normalized field must have a value 
 which is at least two orders of magnitude bigger than the Planck scale. As a result the axion field will evolve very little in the course of the universe's evolution. The slow roll parameter $\epsilon$, eq.(\ref{defe}),
 for example would satisfy the condition
\be
\label{slcond}
\epsilon \lesssim 10^{-4}.
\ee
This  makes the equation of state for quintessence essentially indistinguishable from the cosmological constant. 

In summary, we do not know for certain whether the axion couples to electromagnetism in the form of eq.(\ref{dell}). 
Such a coupling would  arise   if the $U(1)$ gauge field 
 originates on a D5-brane wrapping the same two-cycle from which the axion zero mode also arises. 
In this case the bound on the rotation put by CMB data is very significant. It requires the canonically normalized 
axion field to change very little during the course of the evolution of the universe, thereby making the equation of state 
for   quintessence  essentially like the cosmological constant. In such a situation our best hope for telling
 quintessence apart from the cosmological constant would be to look for a signal  in the  rotation for the CMB 
 polarization itself. 

\subsection{The Linear Potential}
The linear nature of the axion potential is valid for an appropriate range of axion values, we took this form in our discussions above to simplify the analysis. 
 
The $V_0$  contribution in eq.(\ref{pot0}) comes from the induced 3-brane tension and its more accurate form is in 
eq.(\ref{tension}) with $\epsilon = e^{4A_0}$, as noted in \S3.2.
As the axion runs towards its minimum at $a\rightarrow 0$ the  correction  due to the $L^4$ term within the square root
in eq.(\ref{tension}) will become more important.   

Also, if we allow for both negative and positive values of the axion, $a$ in eq.(\ref{pot0}),  eq.(\ref{pot1}) should be replaced by $|a|$. E.g, the more correct form for $V_1$ is     
 \be
\label{pmvalues}
V_1=c_2 M_{SB}^4 e^{2A_0} \left({R^2\over \alpha' L^6}\right) |a|,
\ee
which arises because the change in the volume induced by the warping only depends on $|a|$. 

With $c_2>0$ (which makes physical sense) the total potential  $V_0+V_1$ is positive. Now actually there is an overall 
constant $C$ in the potential, related to the cosmological constant,   which we have not worried about. Including 
this we get the potential to be, 
\be
\label{potreally}
V_{\rm total}=V_0+V_1+C.
\ee

It is worth emphasizing here that the $V_1$ contribution is linear in the axion only in the limit when $|a|\ll N$ where $N$ is the total five -form flux supporting the warped throats where the $5$ branes are placed. Once $|a|$ becomes larger  
there will be corrections and one does not expect the potential to remain linear in the axion, for example the warp factor
$e^{A_0}$ at the bottom of the throat will itself begin to depend on $a$. As was noted in \S4.3 with 
todays' knowledge of warped compactifications we have at best been able to estimate the linear corrections, calculating 
the higher order terms is even harder. For now, we note the possibility that 
including these terms  might well give a more rapidly varying axion potential
for $|a|>N$, and the more rapidly varying potential might actually help improve the tracker behavior of this model,
 to which we turn next. 

\subsection{Tracker Behavior}
It is well known that many quintessence models exhibit tracker  behavior, see e.g., \cite{Peebles:2002gy}. 
The tracker solution is an attractor, and at least for some range of initial conditions the system is drawn to the tracker 
solution regardless of initial conditions. 

Here let us consider the linear potential case, with the net potential being given by eq.(\ref{totpot}). 
The equation of motion for the axion is  given by 
\be
\label{equint}
\ddot{\phi}+3H\dot{\phi}+{\mu^4\over f_a}=0
\ee 
Let us assume  that the universe is radiation dominated, then 
\be
\label{raddom}
H={1\over 2 t}.
\ee
The general   solution to eq.(\ref{equint}) is 
\be
\label{eqtwo}
\phi=c_1-{\mu^4\over 5 f_a}t^2  + c_2 t^{-{1\over 2}}
\ee
where $c_1,c_2$ are  integration constants.  We see that the axion runs to the origin, 
where its energy is minimized in a  time determined by $c_1$. 

Let us  take the solution with $c_2=0$ and some fixed value of $c_1$ and examine perturbations around it. 
There are two kinds of  such perturbations. One is time independent and simply shifts $c_1$, the second dies like 
$t^{-{1\over 2}}$. Since the first perturbation is constant in time we see that this solution is not really a tracker. 

Working instead in  an epoch which is matter dominated sets $H=2/3t$,  the  essential features of the 
axion solution found above continue to be the same in this case as well.   
\subsection{The Cosmological Constant}

The lack of tracker behavior  means that our model does not  solve the coincidence problem. The initial value for the axion, 
 $c_1$,  must be chosen to be just right so energy in the axion field is of the right order of 
magnitude today when the Hubble constant has value $H_0$. 

In fact this feature is tied to the issue of the  cosmological constant in this model which we have mostly ignored so far.
Including the constant $C$ in the total potential, eq.(\ref{potreally}), and  working with the linear potential as an approximation we have,
\be
\label{vtotala}
V_{\rm total}=\mu^4|a|+C
\ee
where we now keep track of the fact that the potential really depends on $|a|$. 

It is now clear that changing the initial conditions for the axion in effect changes $C$. 
Another way to say this is that shifting $a$ and then changing $C$ appropriately can keep $V_{\rm total}$
and in fact the full axion Lagrangian eq.(\ref{finalaxionact})
 invariant \footnote{This is related to some of the discussion
in \S5.3.}.   

An important idea, for which the landscape of string theory has now provided considerable evidence,
 is that the cosmological constant takes its  small value due to anthropic considerations, see, e.g.,
 \cite{Weinberg:1987dv,Barrow:1986,Hogan:1999wh}.  Let us consider how 
such anthropic considerations might work in our model \footnote{In fact anthropic considerations for 
quintessence with a linear potential were studied in \cite{Garriga:1999bf,Garriga:2003hj}. A probability distribution for the 
parameters of this model, partly based on eternal inflation, was assumed in the analysis.}.
Let us take the constant $C$ and the initial condition for the axion $c_1$ as the two variable which can be varied. 
One possibility is that the axion sits at its minimum at $a=0$ and dark energy is entirely due to the  cosmological constant
with $C$ taking the value $\Lambda^4$ due to anthropic considerations. But the more general possibility is that the total energy in dark energy is shared more equitably between the cosmological constant and the axion field with both $C$ and 
$\mu^4 |a|$ being of order $\Lambda^4$. In this latter case, the equation of state for dark energy could show 
significant time variation. In fact, knowing nothing better, one would tend to believe that this latter  possibility is more 
likely, simply because it is more general. Of course deciding this issue more systematically 
 would require a well motivated probability measure in the space of all possibilities, which we lack at the moment. 

Let us emphasize that it was important in the discussion  of this subsection that the axion potential is slowly  varying.  
Instead  suppose the scale in the potential eq.(\ref{totpot}) is,  $\mu^4 \sim M_{SB}^4$, which is  the SUSY breaking scale. 
Then starting with $c_1\sim M_{Pl}$ and taking $f_a\sim M_{Pl}$ for simplicity, we find that  the axion field
runs to its minimum in time $t\sim {M_{Pl}/ M_{SB}^2}$ which is  of order the Hubble constant at the time of 
SUSY breaking. This is too fast to be of  any relevance today. 

\section{Conclusions}
In this paper we have constructed a model for quintessence in string theory.  The model is based on the idea of axion monodromy.
An axion plays the role of the quintessence field, its shift symmetry is broken by the presence of $5$-branes which are located
in highly warped throats. We show that even after including the effects of moduli stabilization and SUSY breaking,
the resulting potential for the axion can be made slowly enough varying to result in a workable model of quintessence. 
If the canonically normalized axion field has an initial expectation value of order the Planck scale, the equation 
of state of dark energy shows significant time variation during the evolution of the universe.  

 Our model has many light particles which arise due to the highly warped throats in which the 5-branes are placed.
The energy scale at the bottom of these throats is at least $10^{-3} \, {\rm eV}$  and typically is much lower. The light particles arise from
 warped down string modes or  KK modes and  are analogous to  the  light particles in the 
Randall Sundrum II model which has only the  Planck brane with  a non-compact extra direction. The couplings of these
  particles to standard model fields and to the four dimensional graviton are highly suppressed, making them difficult to detect.

We have  not attempt to construct a complete model of cosmology, including for example 
inflation at early times.  We have also not attempted to explicitly incorporate 
   the standard model fields  in the kind of construction we used.
This latter issue especially deserves further attention   because  a  coupling of the axion to electromagnetism would 
lead to a  rotation of the
E mode of polarization of the CMB to the B mode, with potential observable consequences. 
We show how such a coupling  can arise  in our model when the photon is the  gauge field   living on  a D5-brane
 and calculate the resulting rotation effect.   

The cosmological constant question is still left open in our discussion.
 Anthropic considerations could well be responsible for its smallness. 
 We suggest that  these  considerations, in the context of our  model, would probably
 lead to the conclusion  that the axion field is not exactly at its minimum. Rather, it is more likely to be  
 rolling and   carrying a reasonable fraction of the total dark energy, thereby causing the  equation of state for 
dark energy to vary. 
Models of the type we present here,  where the potential for the quintessence field
 is slowly enough varying,   might  therefore 
 tilt  the ``betting odds'' in favor of quintessence, even though they lack an 
explanation for the cosmological constant along conventional, as opposed to anthropic, lines.



\section*{Acknowledgment}
We have   benefited from discussions with Kiwoon Choi, Atish Dabholkar, Satoshi Iso, 
 Hideo Kodama, Yoshihisa Kitazawa,
 Shiraz Minwalla,  Yasuhiro Okada, Ashoke Sen, 
 Gary Shiu  and Zheng Sun. We particularly thank Shamit Kachru, Liam McAllister and Eva Silverstein
for their comments. 
YS is grateful to KEK Theory Center for their hospitality, and thanks the organizers and the participants
 of Horiba international conference: COSMO/CosPA 2010, and the IPMU Focus week on cosmology during  
the course of which some of  this work was done.
SPT thanks the organizers and participants of the ExDiP2010 conference where some of these results were presented.
YS and  SPT acknowledge  support from the DAE Govt. of India.
We are especially grateful to the people of India for their sustained support for  research in science.



\providecommand{\href}[2]{#2}\begingroup\raggedright\endgroup

\end{document}